\documentclass[11pt,a4paper]{article}
\usepackage{jheppub}

\usepackage{graphicx}
\usepackage{dcolumn}
\usepackage{bm}
\usepackage{wrapfig}
\usepackage{amsmath,amsfonts,amssymb,bm,ulem}
\usepackage[usenames]{color}
\usepackage{bm}
\usepackage{slashed}
\usepackage{mathrsfs}
\usepackage{mathtools}
\usepackage{graphicx}
\usepackage{dcolumn}
\usepackage{bm}
\usepackage{bbm}
\usepackage{feynmp-auto}

\newcommand{\be}{\begin{equation}}
\newcommand{\ee}{\end{equation}}

\title{Soft Theorems from Boundary Terms in the Classical Point Particle Currents}

\author{Colby DeLisle,}
\emailAdd{cdelisle@phas.ubc.ca}
\author{Jordan Wilson-Gerow,}
\emailAdd{wilsonjs@phas.ubc.ca}
\author{Philip Stamp}
\emailAdd{stamp@phas.ubc.ca}
\affiliation{
	Department of Physics \& Astronomy\\
	and\\
	Pacific Institute for Theoretical Physics\\
	University of British Columbia\\
	Vancouver, BC, V6T 1Z1, Canada
}
\date{\today}

\abstract{Soft factorization has been shown to hold to sub-leading order in QED and to sub-sub-leading order in perturbative quantum gravity, with various loop and non-universal corrections that can be found. Here we show that all terms factorizing at tree level can be uniquely identified as boundary terms that exist already in the classical expressions for the electric current and stress tensor of a point particle. Further, we show that one cannot uniquely identify such boundary terms beyond the sub-leading or sub-sub-leading orders respectively, providing evidence that the factorizability of the tree level soft factor only holds to these orders. Finally, we show that these boundary terms factor out of all tree level amplitudes as expected, in a theory where gravitons couple to a scalar field.}

\begin{document}

\maketitle
\flushbottom

\section{Introduction}

The study of the effect of soft gauge bosons on the dynamics of matter goes back a very long way. In this paper we will be looking at both soft photons and soft gravitons, and focussing on the way in which one can relate not only the leading soft factors, but also the sub-leading and sub-sub-leading factors, to boundary terms in the corresponding classical theory. We begin here by recalling the background to this study, and some of the key issues.

\subsection{Background}

Factorization theorems associated with gauge theories have a long history, beginning with work of Sudakov \cite{sudakov56, abrikosov} for QED; they are implicit in work going back to Schwinger \cite{schwinger49c}. This history is reviewed by Collins \cite{collins89}, and summarized in several texts \cite{sterman,peskinS,weinbergI,LL-QED}. Of interest to us here, for QED, QCD, and quantum gravity, are the ``soft factors'' associated with any scattering involving low-energy gauge bosons. The result that such factors exist is often called the ``soft photon theorem'', or the ``soft graviton theorem,'' depending on the context. For QED the nature of the soft factors is already clear in the work of Murota \cite{murota60} and Yennie et al. \cite{yennie61,yennie2}.

A typical example is that analyzed for both soft photons and soft gravitons in 1965 by Weinberg \cite{weinberg}, in which a set of particles scatter off each other, with the emission of a single graviton or photon. We will be focussing on this particular process for much of this paper - the factorization can be written graphically as

\be \label{eq:pic}
\begin{gathered}
\begin{fmffile}{softgrav1}
  \begin{fmfgraph*}(60,60)
	\fmftop{t1}
	\fmfbottom{b1}
	\fmf{phantom}{b1,v,t1}
	\fmfblob{0.4w}{v}
	\fmfsurroundn{v}{16}
	\fmf{dbl_wiggly, tension=0}{v,v2}
	\fmflabel{$p_1$}{v11}
	\fmflabel{$p'_1$}{v7}
	\fmflabel{$p_n$}{v15}
	\fmflabel{$p'_m$}{v3}
	\fmflabel{$q,\epsilon$}{v2}
	\fmf{plain,tension=0}{v,v3}
	\fmf{plain,tension=0}{v,v7}
	\fmf{plain,tension=0}{v11,v}
	\fmf{plain,tension=0}{v15,v}
	\fmfv{decor.shape=circle,decor.filled=full,decor.size=2thin}{v4,v5,v6,v12,v13,v14}
  \end{fmfgraph*}
\end{fmffile}
\end{gathered}
\hspace{2em}
\approx
\hspace{1em}
\begin{gathered}
\begin{fmffile}{softgrav2}
  \begin{fmfgraph*}(60,60)
	\fmftop{t1}
	\fmfbottom{b1}
	\fmf{phantom}{b1,v,t1}
	\fmfblob{0.4w}{v}
	\fmfsurroundn{v}{16}
	\fmflabel{$p_1$}{v11}
	\fmflabel{$p'_1$}{v7}
	\fmflabel{$p_n$}{v15}
	\fmflabel{$p'_m$}{v3}
	\fmf{plain,tension=0}{v,v3}
	\fmf{plain,tension=0}{v,v7}
	\fmf{plain,tension=0}{v11,v}
	\fmf{plain,tension=0}{v15,v}
	\fmfv{decor.shape=circle,decor.filled=full,decor.size=2thin}{v4,v5,v6,v12,v13,v14}
  \end{fmfgraph*}
\end{fmffile}
\end{gathered}
\times \hspace{-1em}
\begin{gathered}
\begin{fmffile}{softgrav3}
  \begin{fmfgraph*}(60,60)
	\fmftop{t1}
	\fmfbottom{b1}
	\fmf{phantom}{b1,v,t1}
	\fmfsurroundn{v}{16}
	\fmf{dbl_wiggly, tension=0}{v,v2}
	\fmfv{decor.shape=circle,decor.filled=empty,decor.size=0.4w,label=$\hspace{-1.1em}\mathscr{S}$}{v}
	\fmflabel{$q,\epsilon$}{v2}
  \end{fmfgraph*}
\end{fmffile}
\end{gathered}
\ee

Here the graviton momentum $q$ is taken to be very small, and the total soft factor $\mathscr{S}$ depends only on the soft graviton variables $q, \epsilon$. The key point is then that the scattering amplitude picks up a multiplicative factor from the emitted boson.

Because the multiplicative soft factors are divergent in the infrared, they require careful analysis. A key result here, discussed first for QED by Bloch and Nordsieck \cite{blochN}, and thoroughly analyzed  by Kinoshita \cite{kinoshita,kinoshita2}, Nakanishi \cite{nakanishi}, and others (and extended to gravitons by Weinberg \cite{weinberg}), is that divergences due to soft radiation will cancel against divergences due to virtual soft particles. A consistent approach including both divergences then leads to finite cross sections.

Consistent approaches to the IR divergences and factorization can also be set up using a systematic eikonal expansion \cite{fradkin,fradkin2}, which has been employed in recent years to study soft graviton effects in contexts ranging from gravitational decoherence to Hawking radiation \cite{veneziano,veneziano4,GtH87,kabat,giddings,white,veneziano2,veneziano3,us}.  One can also address these divergences by ``dressing'' the asymptotic states of the matter particles using coherent boson states \cite{FK,chung,GR,kibble1,kibble2,kibble3,kibble4,ak1,ak2,kapec1}; these dressed states can be imagined as a ``cloud'' of soft gauge bosons surrounding charged particles.

These divergences (or lack thereof) have many physical effects, which were first discussed in the context of bremsstrahlung by Bloch and Nordsieck \cite{blochN}; they figure in the calculation of radiative processes in many areas of physics. Detailed discussion can be found in many texts \cite{sterman,peskinS,LL-QED,donoghueSM}; the very important applications to astrophysics are surveyed in refs \cite{rybicki04,longair11}. The discussion of radiative corrections in quantum gravity is more recent (see, eg., Donoghue \cite{donoghue12}, and refs. therein).

In recent years many of these ideas have come together in the discussion of several related themes, viz., (i) the black hole information paradox; (ii) asymptotic symmetries in both QED and quantum gravity; and (iii) decoherence caused by the emission of soft photons and gravitons. The black hole information paradox dates back 45 years \cite{hawking-info,hawking-info_1}, and is still very controversial \cite{unruh17,marolf17}. The idea that one might analyze and possibly even resolve this paradox in terms of soft graviton emission, with information stored in the form of asymptotic ``BMS charges'' related to asymptotic symmetries at null boundaries, is however quite recent \cite{strominger,he1,he2,camp7,hawking16,hawking16_1}. In this point of view, the leading soft theorems arise as a consequence of demanding that these asymptotic symmetries are symmetries of the S-matrix. Whether soft gravitons are involved in resolving the black hole information paradox is still unclear.

Since this work, connections have been delineated between the divergent soft factors, in both QED and quantum gravity, and asymptotic symmetries at null infinity (see e.g. \cite{kapec2,camp6,ak3}, and \cite{stromlect} for a review).  The issues involved are subtle; IR divergences have long been known to present complications when trying to directly quantize electromagnetic or gravitational fields asymptotically, at null infinity (see, e.g., Ashtekar \cite{ashtekar1,ashtekar2,ashtekar3}).

The question of how soft gauge bosons cause decoherence of matter states can also be related to asymptotic states. In QED, studies of decoherence are quite old \cite{petruccione,bellomo}; more recent work on both soft photon and soft graviton decoherence  \cite{oniga,us,gordon1} indicates that matter states are almost completely decohered over long times except under very special circumstances, although the problem is not completely resolved \cite{gordon4,gomez}; the connection with asymptotic states can be understood in various ways \cite{us,gordon1}. In this context it is interesting that scattering using dressed states \cite{gordon1} seems to exhibit interference terms which are more in line with what is seen in finite time experiments.

\subsection{Sub-leading and sub-sub-leading terms}

While physically very important, the leading soft theorem and its physical applications are not the whole story. It has been understood for some time that there are sub-leading corrections \cite{low1,low2,BK,GG,delduca,jackiw1,jackiw2}. Further connections between asymptotic symmetries and these sub-leading corrections to the soft theorems have also been explored \cite{lysov,camp1,camp2,conde1,camp3,camp4,camp5,conde2,laddha2}, in hopes that the asymptotic structure which explains the leading order might also account for the sub-leading ones.

Of particular relevance to us is work in the last few years investigating sub-leading terms \cite{soft,white,loops1,loops2}, in which amongst other things it is conjectured that there are specific sub-leading and sub-sub-leading corrections to the soft graviton theorem. This was shown to be true for perturbative gravitational amplitudes at tree level; loop corrections were discussed in \cite{loops1,loops2}, and \cite{sahoo1} derives loop corrected results independent of any IR cutoff. Consider a tree level amplitude $A$, and the same amplitude but with an extra soft graviton with momentum $q$ and polarization $\epsilon_{ab}$ in the out state, $\mathscr{A}( q, \epsilon)$. The claim is then that, \textit{at tree level},
\be
\mathscr{A}( q, \epsilon) \;\;=\;\;   \left(  \mathscr{S}_{(0)} + \mathscr{S}_{(1)} + \mathscr{S}_{(2)} \right) A+ \mathcal{O}(|\mathbf{q}|^2).
\ee
as $|\mathbf{q}|\rightarrow 0$, with $\mathscr{S}_{(k)} \propto |\mathbf{q}|^{k-1}$.

The first term $ \mathscr{S}_{(0)}$ is the usual divergent leading soft factor. The second term $ \mathscr{S}_{(1)}$ was motivated by proposed extensions to the gravitational symmetries of null infinity. The sub-sub-leading term $ \mathscr{S}_{(2)}$ is somewhat more enigmatic. Its existence has been proven at tree level, but although it seems clear that the soft theorems are intimately related to asymptotic symmetries, there is currently no conclusive evidence that such a description explains the appearance of $\mathscr{S}_{(2)}$.

In this work, we propose an explanation for the appearance of this term. As we will see, there is something special about the terms which factorize at tree level --- the leading, sub-leading, and sub-sub-leading soft graviton factors.  Drawing inspiration from eq. \eqref{eq:pic}, we expect that in order to factorize, the total soft factor $\mathscr{S}$ can depend only on (i) the soft graviton variables $q, \epsilon$, (ii) the initial hard momenta $\{p_n\}$, and (iii) the final hard momenta $\{p'_m\}$ (or derivatives with respect to these). In other words, we expect that the soft factor can only involve what the graviton field can interact with at the infinite future and past endpoints of the particle worldlines in the diagram.

Here we will make this intuition precise. By looking at a particular soft limit of the source of gravitational radiation from a massive point particle --- the Fourier transform of the particle's stress tensor $T^{ab}$ --- we will find that only certain contributions to the \textit{classical} stress tensor can be written as boundary terms which localize to the endpoints of the particle's worldline. We will show that the \textit{only} such contributions to $T^{ab}$ are simply related to the leading, sub-leading, and sub-sub-leading soft graviton factors. This result will be made precise in eq. \eqref{eq:Tbdry} below.

It is remarkable that such a simple criterion singles out exactly these terms, while predicting no further terms of the same type. Indeed, we will also show that a similar manipulation of the electromagnetic current $j^a$ produces precisely the leading and sub-leading soft photon factors. This will lead to the result \eqref{eq:Jbdry}. The equations \eqref{eq:Tbdry} and \eqref{eq:Jbdry} are our main results.

The idea of relating the Fourier transform of the particle sources and resulting radiative fields to the soft theorems is not new. Particularly relevant is work that has been done on the so-called ``classical soft theorems.'' For example in \cite{laddha0.1}, the classical limit of the soft graviton theorem was taken up to sub-sub-leading order, in order to predict the classical radiation fields at early and late retarded times produced by scattering events. This prediction was then verified directly in the classical theory in five or more dimensions \cite{laddha0.2} and in four dimensions \cite{laddha1,saha,sahoo2}. Interestingly, these latter works also identify logarithmic corrections to the soft theorems in four dimensions which arise beyond leading order because of long range forces between the particles in the in and out states asymptotically. 

Our work seeks to build on this connection between soft theorems and classical radiation at early and late times. Specifically, here we relate the soft theorems directly to the matter currents by looking for boundary terms in those currents, rather than explicitly taking the limits $t\rightarrow\pm\infty$ in the resulting radiation field. This identification of unambiguous boundary terms is new, and explains the factorizability of the tree level soft factors. Our proof that there are no new terms of this type at higher orders is also novel, and provides evidence against further factorization at higher orders.

The structure of the paper is as follows. In section \ref{sec:prelim} we review the soft photon and soft graviton theorems, and establish our notation. Section \ref{sec:EM} then contains the manipulations of the classical electromagnetic current $j^a$ which isolate the leading and sub-leading soft photon factors. In section \ref{sec:gravity} we perform the same manipulations on $T^{ab}$ to yield the leading, sub-leading, and sub-sub-leading soft graviton factors. This section also highlights one major difference in the derivation between the electromagnetic and gravitational cases. We then conclude in section \ref{sec:done}. In the Appendix \ref{app.b} of the paper we give a detailed demonstration of how our results show that the three soft graviton terms factorize out of all tree level scattering amplitudes, in a simple model containing only gravitons and a scalar field.

In this work, we use the mostly-minus metric signature, $\eta_{ab} = \textrm{diag}(+1, -1, -1, -1)$. Vectors with only spatial components will be bolded, otherwise all vectors have four space-time indices. An overbar on a quantity denotes its complex conjugate. Lastly, a dot on a quantity here will always mean a derivative with respect to proper time, $\dot{X} \equiv \frac{d}{ds}X$.

\section{Soft limit preliminaries} \label{sec:prelim}

We begin by looking at the case mentioned in the introduction, i.e., with the amplitude for a process in which a set of $N$ matter particles with momenta $\{p_n\}$ scatter into a set of $M$ particles with momenta $\{p_m'\}$. The subscripts $n$ and $m$ label each particle in the in/out states. We will call this amplitude $A(\{p_m'\}|\{p_n\})$.

Now consider the amplitude for the same process, but add to the out state a photon with momentum $q$ and polarization vector $\epsilon$. We call this modified amplitude $\mathcal{A}(\{p_m'\};q, \epsilon|\{p_n\})$, and
one finds that, at tree level \cite{low1,low2,BK,GG}:
\be \label{eq:SPT}
 \mathcal{A}(\{p_m'\}; q, \epsilon|\{p_n\}) \;\;= \;\; \left(  \mathcal{S}_{(0)} + \mathcal{S}_{(1)} \right) A(\{p_m'\}|\{p_n\}) + \mathcal{O}(|\bf{q}|),
\ee
as $|\mathbf{q}|\rightarrow 0$, with $\mathcal{S}_{(k)} \propto |\mathbf{q}|^{k-1}$. More concretely, we have
\be \label{eq:softQED}
\mathcal{S}_{(0,1)} \;=\;  \mathcal{N}_a \left[\sum_{m=1}^M e_m \mathcal{S}_{(0,1)}^a(q,\hat{x'}_m,p'_m) - \sum_{n=1}^N e_n \mathcal{S}_{(0,1)}^a(q,\hat{x}_n,p_n) \right],
\ee
where $\mathcal{N}_a$ is the ``wavefunction'' of the outgoing photon, i.e.,
\be
\mathcal{N}_a \equiv  \frac{\bar{\epsilon}_a}{(2\pi)^{3}\sqrt{2|\mathbf{q}|}} ,
\ee
where the $e_{m,n}$ are the charges of the matter particles, and where we have defined
\begin{align}
\mathcal{S}_{(0)}^a(q,p) &\equiv   \frac{p^a}{q\cdot p} \\
\mathcal{S}_{(1)}^a(q,\hat{x},p) &\equiv  i \frac{q_b\hat{J}^{ba}}{q\cdot p} .
\end{align}
Here $\hat{J}^{ab} \equiv p^a \hat{x}^b - \hat{x}^a p^b \equiv 2 p^{[a} \hat{x}^{b]}$ is the angular momentum operator, in which the position operator is represented by derivatives with respect to momentum when working in the momentum eigenbasis.

Instead of the photon, we can also add to the out state a graviton with momentum $q$ and polarization tensor $\epsilon$. We denote this graviton-emission amplitude by $\mathscr{A}(\{p_m'\};q, \epsilon|\{p_n\})$.
One now finds that, at tree level:
\be \label{eq:SGT}
\mathscr{A}(\{p_m'\}; q, \epsilon|\{p_n\}) \;\;=\;\;  \left(  \mathscr{S}_{(0)} + \mathscr{S}_{(1)} + \mathscr{S}_{(2)} \right) A(\{p_m'\}|\{p_n\}) + \mathcal{O}(|\mathbf{q}|^2)
\ee
near $|\mathbf{q}| = 0$, where once more
\be
\mathscr{S}_{(0,1,2)} \;=\;  \kappa \mathscr{N}_{ab} \left[\sum_{m=1}^M  \mathscr{S}_{(0,1,2)}^{ab}(q,\hat{x'}_m,p'_m) - \sum_{n=1}^N  \mathscr{S}_{(0,1,2)}^{ab}(q,\hat{x}_n,p_n) \right],
\ee
in which $\kappa \equiv \sqrt{8\pi G}$ is the Newton coupling, $\mathscr{N}_{ab}$ the outgoing graviton ``wavefunction," viz.,
\be \label{eq:Ngrav}
\mathscr{N}_{ab} \equiv  \frac{\bar{\epsilon}_{ab}}{(2\pi)^{3}\sqrt{2|\mathbf{q}|}} ,
\ee
and
\begin{align}
\mathscr{S}_{(0)}^{ab}(q,p) &\equiv   \frac{p^ap^b}{q\cdot p} \\
\mathscr{S}_{(1)}^{ab}(q,\hat{x},p) &\equiv  i \frac{q_c \hat{J}^{c(a}p^{b)}}{q\cdot p} \\
\mathscr{S}_{(2)}^{ab}(q,\hat{x},p) &\equiv  -\frac{1}{2} \frac{q_c\hat{J}^{ac} q_d \hat{J}^{bd}}{q\cdot p} ,
\end{align}
where the parentheses around indices indicate symmetrization on those indices.

Equations \eqref{eq:SPT} and \eqref{eq:SGT} are the tree level soft photon theorem and soft graviton theorem, to sub-leading and sub-sub-leading order respectively. In what follows we will show how each of these factors is already encoded in boundary terms in the classical point particle electromagnetic current and stress tensor. Our reasoning also explains why there are in general no further factorizing terms at higher orders.

\section{Electrodynamics} \label{sec:EM}

In this section, we will look at the conserved electromagnetic current associated with a moving classical point particle. We first demonstrate that the soft factors appearing in the leading and sub-leading soft photon theorems are already identifiable within the Fourier transform of the current at small energies. After finding these, we will show explicitly that there are no further terms of this type in the small energy expansion of the current.

\subsection{Finding the soft factors in the classical current}

At the level of the action, a charged particle couples to the electromagnetic field through the interaction term
\be
S_{int} \;=\;  \int d^4x A_a(x) j^a(x) \;=\;   \int \frac{d^4q}{(2\pi)^4} A_a(q) j^a(-q)
\ee
in which $A_a$ can be thought of as representing a ``soft photon'' if we take the ``photon'' momentum $q^a$ to be on-shell ($q\cdot q = 0$), and of very low frequency. The matter then couples to the photon field via
\be
j^a(-q),
\ee
where we write $q^a = |\mathbf{q}|(1, \hat{n})$, where $\hat{n}$ is a spatial unit vector, and $|\mathbf{q}|$ is taken to be small, $|\mathbf{q}| \rightarrow 0$. The Fourier transform of the current evaluated on a null four-momentum with vanishingly small frequency is thus the object of our study, since this is what interacts directly with on-shell soft photons.

The conserved current for a particle with charge $e$ (and mass $m$) coupled to the Maxwell field, which follows a classical trajectory $X^a(s)$, is
\be \label{eq:EMcurrent}
j^a(x) \;=\;  e \int_0^\infty ds\,\dot{X}^a(s) \, \delta^{(4)}(x - X(s)),
\ee
in which $s$ is the particle's proper time, and the overdot represents differentiation with respect to this proper time, i.e., $\dot{X}(s) \equiv \frac{d}{ds}X(s)$. The integral is taken over the entire trajectory of the particle, which we take to extend infinitely far in the past and future directions. In the infinite past, no proper time has yet elapsed ($s=0$), and it takes an infinite amount of time to reach the infinite future $s\rightarrow \infty$.

Looking at the Fourier transform of this, $j^a(q) \equiv \int d^4x \, e^{iq\cdot x} j^a(x)$, we get
\be
j^a(q) \;=\;  e \int_0^\infty ds\,\dot{X}^a(s) \, e^{iq\cdot X(s)} .
\ee
This will be the starting point for our search for the soft factors; from here on we will suppress the dependence of $X^a$ on $s$, as well as the integration limits.

Let us first rewrite this current using a simple identical transformation, i.e.,
\be
j^a(q) \;=\;  e \int ds\, \dot{X}^a \left( \frac{1}{iq\cdot \dot{X}} \right) \frac{d}{ds} e^{iq\cdot X}.
\ee
so that integrating by parts in $s$ gives
\be \label{eq:with_bdry_term}
j^a(q) \;\;=\;\;  -ie \int ds\,\frac{d}{ds} \left(  e^{iq\cdot X}  \frac{\dot{X}^a}{q\cdot \dot{X}}  \right) + ie \int ds\, e^{iq\cdot X} \frac{d}{ds} \left( \frac{\dot{X}^a}{q\cdot \dot{X}} \right) .
\ee

In what follows we will concern ourselves with scattering processes, in which case the limits of integration correspond to the infinite past and infinite future (as seen pictorially in \eqref{eq:pic}). The boundary term in this expression is thus ill-behaved and should be dropped, since it will be proportional to the oscillatory factor $\exp(\pm iq^{0}\infty)$. More specifically, the term in the infinite future is proportional to $\exp(i q^0 \infty)$, and so vanishes if we take the prescription $q^0 \rightarrow q^0 +i\epsilon$, while the past term vanishes when taking $q^0 \rightarrow q^0 -i\epsilon$. Dropping the boundary term then renders the expression well-behaved, as the second term is only nonzero when $\ddot{X} \neq 0$, which is not the case asymptotically giving the integrand a finite limit in the infinite future and past.

This point has been raised before in \cite{ilderton}, which agrees with our $i\epsilon$ prescriptions, and also gives some physical intuition for the problematic boundary term. The current \eqref{eq:EMcurrent} is identically conserved, $j^a(x)_{,a} = 0$. In momentum space, we should have that $q_a j^a(q) = 0$, but contracting $q_a$ with \eqref{eq:with_bdry_term} yields a generally nonzero value coming only from the boundary term:
\be
q_a j^a(q) = -ie \int ds\,\frac{d}{ds} \left(  e^{iq\cdot X}  \right) \neq 0
\ee
Enforcing current conservation gives us yet another justification for dropping the boundary term, and from this point forward we will always neglect such terms. Similar conclusions about this kind of integral have been drawn in \cite{laddha1}, and a careful discussion of the $i\epsilon$ prescription turns out to be especially important when discussing logarithmic corrections to the soft theorems \cite{sahoo2,saha}.

For us, the important thing to remember is that we should drop the oscillatory boundary term, leaving us with the following expression for the particle current:
\be
j^a(q) =  ie \int ds\, e^{iq\cdot X} \frac{d}{ds} \left( \frac{\dot{X}^a}{q\cdot \dot{X}} \right)
\ee
From here we want to look at what happens to this expression in the limit of small frequency. Near $|\mathbf{q}|=0$ we assume we can Taylor expand the phase in the integrand (where we now set $q^a = |\mathbf{q}|(1, \hat{n})$). We then have
\be
j^a(q) \;=\;  ie \int ds\, \left[  \sum_{k=0}^\infty \frac{1}{k!}(iq\cdot X)^k \right] \frac{d}{ds} \left( \frac{\dot{X}^a}{q\cdot \dot{X}} \right)
\equiv  \sum_{k=0}^\infty j_{(k)}^a(q),
\ee
which defines the $k$th order current $j_{(n)}$. The leading order term is
\be
j_{(0)}^a(q) \;=\;   ie \int ds\,\frac{d}{ds} \left( \frac{\dot{X}^a}{q\cdot \dot{X}} \right)  = ie \Delta \left( \frac{\dot{X}^a}{q\cdot \dot{X}} \right),
\ee
where we have introduced the symbol $\Delta$ to stand for the difference between infinite future and infinite past values, i.e., $\Delta f(s) \equiv f(s\rightarrow\infty) - f(s\rightarrow-\infty)$. We now see that the zeroth order current is essentially the leading soft factor (with the momentum $P$ being replaced by $m\dot{X}$), i.e. we can write
\be
ij_{(0)}^a(-q) \;=\;  e\Delta \mathcal{S}^a_{(0)}(q, m\dot{X}).
\ee
The $\Delta$ here even explains the relative minus sign between outgoing and incoming particles in \eqref{eq:softQED} --- it comes from the fact that the soft factor is a boundary term evaluated on future/past boundaries.

Going now to first order, we have
\begin{align}
j_{(1)}^a(q) \;&=\;  ie \int ds\, (iq\cdot X) \frac{d}{ds} \left( \frac{\dot{X}^a}{q\cdot \dot{X}} \right)  \nonumber\\
&=\; -e \int ds\, \left[ \frac{d}{ds} \left( q\cdot X \frac{\dot{X}^a}{q\cdot \dot{X}} \right)  - (q\cdot \dot{X}) \frac{\dot{X}^a}{q\cdot \dot{X}}  \right]\nonumber \\
 &=\; -e \int ds\,\frac{d}{ds} \left( q\cdot X \frac{\dot{X}^a}{q\cdot \dot{X}}  - X^a \right),
\end{align}
and we see again that this order too is simply a boundary term. Now, we can re-multiply the $X^a$ in the second term by one, $1 = (q\cdot\dot{X})/(q\cdot\dot{X})$, and rewrite this as
\be
j_{(1)}^a(q) \;=\;  -e \int ds\,\frac{d}{ds} \left( q_b \frac{\dot{X}^aX^b - X^a \dot{X}^b}{q\cdot \dot{X}} \right) \;\;=\;\;  e \Delta \left( \frac{q_b J^{ba}}{q\cdot P} \right),
\ee
where in the second equality we've multiplied by $1 = m/m$ to convert $\dot{X}$ to $P$, and we see the emergence of the orbital angular momentum $J^{ab}$. Clearly this is purely a boundary term, and has the same relationship to the sub-leading soft factor as the leading order current had to the leading soft factor:
\be
ij_{(1)}^a(-q) \;=\;  e\Delta \mathcal{S}^a_{(1)}(q, X, m\dot{X})
\ee

So what we have now shown is that, if we Taylor expand for small $|\mathbf{q}|$, the current for a point particle takes the form
\be \label{eq:Jbdry}
ij^a(-q) \;=\;  e \left[ \Delta \mathcal{S}^a_{(0)}(q, m\dot{X}) + \Delta \mathcal{S}^a_{(1)}(q, X, m\dot{X}) \right] + \mathcal{O}(|\mathbf{q}|),
\ee
where the first two terms are localized to the boundary of the particle's worldline. This demonstrates explicitly how the soft factors are connected to the boundary contribution to the particle current, and is the one of the main results of the paper.

\subsection{Proving there are no further unambiguous boundary terms}

We have shown that the leading and sub-leading soft factors are present in the classical particle current. Can we use this method to obtain further soft factors? The answer will turn out to be no. At the next (sub-sub-leading) order, we have
\begin{align}
j_{(2)}^a(q) \;&=\;  \frac{ie}{2} \int ds\, (iq\cdot X)^2 \frac{d}{ds} \left( \frac{\dot{X}^a}{q\cdot \dot{X}} \right)  \nonumber\\
&=\; \frac{-ie}{2} \int ds\, \left[ \frac{d}{ds} \left( (q\cdot X)^2  \frac{\dot{X}^a}{q\cdot \dot{X}} \right)  - \frac{d}{ds}\left( (q\cdot X)^2 \right)\frac{\dot{X}^a}{q\cdot \dot{X}}  \right].
\end{align}
The first term again is a boundary contribution, but this time, it is not possible to write the second term in the integrand as a total derivative in $s$. The term in question is
\be
\frac{d}{ds}\left( (q\cdot X)^2 \right)\frac{\dot{X}^a}{q\cdot \dot{X}} \;\;=\;\; 2(q\cdot X)(q\cdot\dot{X})\frac{\dot{X}^a}{q\cdot \dot{X}} \;\;=\;\; 2(q\cdot X) \dot{X}^a
\ee
and clearly this isn't just a total derivative. There is thus no \textit{unique} boundary term at this order to be identified as one of the soft factors. Indeed, we see that at this order, we can always perform integrations by parts to move various terms to the boundary, but we are always left with some ``bulk'' contribution, that doesn't just depend on the boundary data. A calculation similar to that in appendix \ref{app.b} shows that the unambiguous identification of terms which only depend on boundary data, at leading and sub-leading orders, is what allows those terms to factor out of the (tree level) amplitudes which include soft photons.

It is now straightforward to show that the breakdown at sub-sub-leading order actually persists to all higher orders in $|\mathbf{q}|$. We have that
\begin{align}
j_{(k)}^a(q) \;\;&=\;\;  \frac{ie}{k!} \int ds\, (iq\cdot X)^k \frac{d}{ds} \left( \frac{\dot{X}^a}{q\cdot \dot{X}} \right)  \nonumber\\
&=\;\; \frac{(i^{k+1})e}{k!} \int ds\, \left[ \frac{d}{ds} \left( (q\cdot X)^k  \frac{\dot{X}^a}{q\cdot \dot{X}} \right)  - \frac{d}{ds}\left( (q\cdot X)^k \right)\frac{\dot{X}^a}{q\cdot \dot{X}}  \right].
\end{align}
Yet again, the first term is already a boundary term. The expression appearing in the second term of the integrand is
\begin{align} \label{eq:jerror}
\frac{d}{ds}\left( (q\cdot X)^k \right)\frac{\dot{X}^a}{q\cdot \dot{X}} \;\;&=\;\; k(q\cdot X)^{k-1} \dot{X}^a  \nonumber\\
&=\;\; \frac{d}{ds} \left( k(q\cdot X)^{k-1} X^a \right) - k(k-1)(q\cdot X)^{k-2}(q\cdot \dot{X})X^a .
\end{align}

From the first equality, it is clear that this term isn't a total derivative. In the second equality, we have written the term as a total derivative, less a term that explicitly vanishes when $k = 0,1$. The second term in the last expression is thus a sort of measure of the failure of each term to localize to the boundary, and it vanishes only at leading and sub-leading orders.


\section{Gravity} \label{sec:gravity}

We turn now to the classical stress tensor of a point particle, via which our particle couples to the metric perturbation in linearized gravity. In this discussion, we will find that our method for obtaining soft factors tracks the procedure used in the previous section rather closely.

There are however two major differences. Most obviously, we will find three unambiguous boundary terms in the stress tensor, instead of the two boundary terms found in the electromagnetic current. The second and more subtle difference is that in what follows, we find ``bulk'' corrections to the boundary terms at sub-leading and sub-sub-leading orders --- at these orders, we do not get precisely the boundary terms we are looking for. However, these bulk corrections vanish when the equations of motion are satisfied.

\subsection{The soft factors in the classical stress tensor: leading term}

The gravitational current for a particle with charge mass $m$ following trajectory $X^a(s)$ is the stress tensor
\be \label{eq:Gcurrent}
T^{ab}(x) \;=\; m \int ds\,\dot{X}^a(s)\dot{X}^b(s) \, \delta^{(4)}(x - X(s))
\ee
and, as in the electromagnetic case, we are interested in the Fourier transform of this:
\be
T^{ab}(q) \;=\; m \int ds\,\dot{X}^a(s)\dot{X}^b(s) \, e^{iq\cdot X(s)}
\ee

Just as before, we can ``multiply by one'' with an insertion of $(1/iq\cdot \dot{X}) \frac{d}{ds}$ acting on the exponential. We then integrate by parts, and drop the terms proportional to $\exp(\pm iq^{0}\infty)$, leaving us with
\be
T^{ab}(q) \;=\;   im \int ds\, e^{iq\cdot X} \frac{d}{ds} \left( \frac{\dot{X}^a\dot{X}^b}{q\cdot \dot{X}} \right) .
\ee

Near $|\mathbf{q}|=0$ (where once again we set $q^a = |\mathbf{q}|(1, \hat{n})$) we can expand this expression in powers of $q$, as
\be
T^{ab}(q) \;=\;   im \int ds\, \left[  \sum_{k=0}^\infty \frac{1}{k!}(iq\cdot X)^k \right] \frac{d}{ds} \left( \frac{\dot{X}^a\dot{X}^b}{q\cdot \dot{X}} \right)
\;\; \equiv \;\;   \sum_{k=0}^\infty T_{(k)}^{ab}(q),
\ee
which defines the $k$th order term $T_{(k)}$ in the stress tensor. The leading order term is again straightforward, and we have
\be
T_{(0)}^{ab}(q) \;=\;   im \int ds\,\frac{d}{ds} \left( \frac{\dot{X}^a\dot{X}^b}{q\cdot \dot{X}} \right)  \;\;=\;\;  im \Delta \left( \frac{\dot{X}^a\dot{X}^b}{q\cdot \dot{X}} \right).
\ee
This clearly agrees with the zeroth order soft graviton factor, viz.,
\be
iT_{(0)}^{ab}(-q) = \Delta \mathscr{S}^{ab}_{(0)}(q, m\dot{X}).
\ee
verifying that we recover the leading soft factor from the lowest term.

\subsection{Sub-leading term: the acceleration terms}

Turning to the first order stress tensor, we immediately find that things are different from the case of electrodynamics. We have
\begin{align}
T_{(1)}^{ab}(q) \;&=\;  im \int ds\, (iq\cdot X) \frac{d}{ds} \left( \frac{\dot{X}^a\dot{X}^b}{q\cdot \dot{X}} \right)  \nonumber\\
&=\; -m \int ds\, \left[ \frac{d}{ds} \left( q\cdot X \frac{\dot{X}^a\dot{X}^b}{q\cdot \dot{X}} \right)  - (q\cdot \dot{X}) \frac{\dot{X}^a\dot{X}^b}{q\cdot \dot{X}}  \right] \nonumber \\
&=\; -m \int ds\, \left[ \frac{d}{ds} \left( q\cdot X \frac{\dot{X}^a\dot{X}^b}{q\cdot \dot{X}} \right)  -\dot{X}^a\dot{X}^b  \right]
\end{align}
and unfortunately the last term is no longer a boundary term!

However things are not as bad as they appear. If we express the last term in the integrand as
\be
\dot{X}^a\dot{X}^b \;\;=\;\;  \frac{d}{ds} \left(  X^{(a}\dot{X}^{b)}   \right)   -   X^{(a}\ddot{X}^{b)},
\ee
where we have explicitly symmetrized on $a$ and $b$, then we can express the sub-leading stress tensor as
\be
T_{(1)}^{ab}(q) \;=\; -m \int ds\, \left[ \frac{d}{ds} \left( q\cdot X \frac{\dot{X}^a\dot{X}^b}{q\cdot \dot{X}}  - X^{(a}\dot{X}^{b)}  \right)  +X^{(a}\ddot{X}^{b)}  \right] ,
\ee
or, rearranging and again using $m$ to turn $\dot{X}$ into $P$,
\begin{align} \label{eq:T_1}
T_{(1)}^{ab}(q) \;\;&=\;\; -m \int ds\, \left[ \frac{d}{ds} \left( q_c \frac{\dot{X}^a\dot{X}^b X^c - X^{(a}\dot{X}^{b)} \dot{X}^c }{q\cdot \dot{X}}  \right)  +X^{(a}\ddot{X}^{b)}  \right]  \nonumber \\
&=\;\;  \int ds\, \left[ \frac{d}{ds} \left( \frac{q_c J^{c(a} P^{b)}}{q\cdot P}  \right)   - mX^{(a}\ddot{X}^{b)}  \right] \nonumber \\
&=\;\;  \Delta \left( \frac{q_c J^{c(a} P^{b)} }{q\cdot P}  \right)  -m \int ds \,X^{(a}\ddot{X}^{b)}    .
\end{align}
This is very close to including only the soft factor $\Delta\mathscr{S}^a_{(1)}$; but we also have an extra ``acceleration"  contribution $-m \int ds \,X^{(a}\ddot{X}^{b)}$, and it is not immediately clear what to do with this.

To understand terms of this form, linear in the particle's acceleration, we consider the (flat space-time) divergence of the stress tensor, viz.,
\be
\begin{split}
T^{ab}_{,a}(x) \;&=\;  m \int ds \, \frac{d}{dx^a}\biggl[ \frac{dx^a}{ds} \frac{dx^b}{ds} \delta^{(4)}( x - X(s))   \biggr] \\
&=\; m \int ds \, \biggl[ \frac{dx^a}{ds} \frac{dx^b}{ds} \biggl(\frac{d}{dx^a} \delta^{(4)}(x - X(s))  \biggr)  + (\textrm{terms like }\frac{d}{dx^a}\frac{dx^b}{ds}) \biggr]
\end{split}
\ee
where the terms not written explicitly vanish as
\be
\frac{d}{dx^a}\frac{dx^b}{ds} \;=\;  \frac{d}{ds}\frac{dx^b}{dx^a} \;=\;  \frac{d}{ds} \delta^b_a \;\;=\;\; 0 .
\ee
Further, using $(dx^a/ds)(d/dx^a) = d/ds$, we have
\be
{T^{ab}}_{,a} \;=\;  m \int ds \,  \frac{dx^b}{ds} \biggl(\frac{d}{ds} \delta^{(4)}( x -  X(s))  \biggr)  ,
\ee
which we can integrate by parts to get (after dropping another boundary term which will turn out to be proportional to $\exp(\pm iq^{0}\infty)$)
\be
{T^{ab}}_{,a} \;=\;  - m \int ds \,  \frac{d^2x^b}{ds^2} \delta^{(4)}( x - X(s)) \;\;=\;\; - m \int ds \,  \ddot{X}^b \delta^{(4)}( x - X(s)).
\ee

Finally, to make contact with our low energy expansion, we can take the Fourier transform of this to get
\be \label{eq:divq}
T^{ab}_{,a}(q) \;=\;  - m \int ds \,  e^{iq\cdot X} \ddot{X}^b .
\ee

Now, from the linearized Einstein equations we know that for the matter stress tensor, ${T^{ab}}_{,a} = 0$. Thus we also have $T^{ab}_{,a}(q)= 0 $ when the equations of motion are satisfied. Expanding the phase in \eqref{eq:divq} gives
\be \label{eq:divk}
T^{ab}_{,a}(q) \;=\;  - m \int ds \,  \left[ \sum_{k=0}^\infty \frac{1}{k!} (i q\cdot X)^k   \right] \ddot{X}^b ,
\ee
and when stress-energy is conserved, these terms must vanish independently for each $k$:
\be
0 = - m \int ds \,  \left[ \frac{1}{k!} (i q\cdot X)^k   \right] \ddot{X}^b
\ee

At lowest order this implies conservation of momentum over the whole of the particle's worldline,
\be
0 \;\;=\;\;  - m \int ds \,  \ddot{X}^b ,
\ee
and we also have the sub-leading and sub-sub-leading orders which will be useful for us,
\begin{align}
0 \;\;&=\;\; - im q_a \int ds \, X^{a} \ddot{X}^{b}   \\
0 \;\;&=\;\; \frac{m}{2} q_c q_a \int ds \,  X^{c} X^{a} \ddot{X}^{b} .
\end{align}

Actually, because $q_a$ here is arbitrary, we have the slightly stronger result
\begin{align}
0 \;\;&=\;\; - im \int ds \, X^{a} \ddot{X}^{b}   \label{eq:divsub} \\
0 \;\;&=\;\; \frac{m}{2}  \int ds \,  X^{c} X^{a} \ddot{X}^{b} , \label{eq:divsubsub}
\end{align}
and we see from \eqref{eq:divsub} that the problematic term in \eqref{eq:T_1} vanishes when the stress tensor is conserved, which it is when the equations of motion are satisfied. This analysis extends trivially to the stress tensors of multiple point particles -- the analog of eqs. \eqref{eq:divsub} and \eqref{eq:divsubsub} will simply contain a sum of terms of the same form, which vanishes on-shell. This allows us to establish that
\be
iT_{(1)}^{ab}(-q) \;\;=\;\; \Delta\mathscr{S}^{ab}_{(1)}(q, X, m\dot{X}) + (\textrm{terms that vanish when }{T^{ab}}_{,a} = 0),
\ee
and although we didn't obtain an expression purely localized at the boundaries of the worldline, what we have here is sufficient to understand the sub-leading soft graviton theorem, as mentioned in appendix \ref{app.b}.

It is important to understand that in linearized gravity, conservation of the stress-energy necessarily fails to include contributions from the stress-energy of the gravitational field. What this means is that these equations we see here predict the dynamics for point particles which are free except for contact interactions with one another -- i.e. they are not subject to long range gravitational forces. One needs to go beyond the linearized limit to incorporate gravitational effects in the stress tensor. Again this will not end up affecting a tree level analysis.

We note that we did not need to have a discussion like this one in the electromagnetic case. This is essentially because the electromagnetic current is \textit{identically} conserved, rather than only being conserved when the field equations are obeyed.

\subsection{Sub-sub-leading order: back to the classical stress tensor}

Let us now consider the next term (of sub-sub-leading order) in $|\mathbf{q}|$. We start with
\be
T^{ab}_{(2)}(q) \;=\;  \frac{im}{2} \int ds\,  (iq\cdot X)^2  \frac{d}{ds} \left( \frac{\dot{X}^a\dot{X}^b}{q\cdot \dot{X}} \right),
\ee
and use the same integration by parts trick as before; this gives
\be
T^{ab}_{(2)}(q) \;=\;  -\frac{im}{2} \int ds\, \left[    \frac{d}{ds} \left( (q\cdot X)^2 \frac{\dot{X}^a\dot{X}^b}{q\cdot \dot{X}} \right)   -      2  (q\cdot X)  \dot{X}^a\dot{X}^b  \right].
\ee
The second term here can be rewritten as
\be
-2  (q\cdot X)  \dot{X}^a\dot{X}^b \;=\; \frac{d}{ds} \left( (q\cdot \dot{X})X^a X^b - 2 (q\cdot X) X^{(a}\dot{X}^{b)}   \right) - X^a X^b (q\cdot \ddot{X}) + 2 (q\cdot X) X^{(a}\ddot{X}^{b)}
\ee
(by setting $k=2$ in eq. \eqref{eq:proven_in_app}), leaving us with
\be
\begin{split}
T^{ab}_{(2)}(q) \;\;=\;\;  -&\frac{im}{2} \int ds\,  \frac{d}{ds} \left( (q\cdot X)^2 \frac{\dot{X}^a\dot{X}^b}{q\cdot \dot{X}} +  (q\cdot \dot{X})X^a X^b - 2 (q\cdot X) X^{(a}\dot{X}^{b)} \right)   \\
 & \qquad\qquad +\; \frac{im}{2} \int ds\, \left[ X^a X^b (q\cdot \ddot{X}) + 2 (q\cdot X) X^{(a}\ddot{X}^{b)} \right] .
\end{split}
\ee

The first line of this equation is a boundary term, and each term in the second line vanishes after invoking the equations of motion, thanks to eq. \eqref{eq:divsubsub}. As expected, we can now rewrite this in a way that is suggestive of the sub-sub-leading soft factor. We get
\be
\begin{split}
T^{ab}_{(2)}(q) \;\;=\;\;  -&\frac{im}{2} \int ds\,  \frac{d}{ds} \left( 4\frac{q_c  (\dot{X}^{[a}X^{c]} )q_d(\dot{X}^{[b}X^{d]} )}{q\cdot \dot{X}} \right)   \\
 & \qquad\qquad +\; \frac{im}{2} \int ds\, \left[ X^a X^b (q\cdot \ddot{X}) + 2 (q\cdot X) X^{(a}\ddot{X}^{b)} \right] ,
\end{split}
\ee
which after using $m\dot{X} \rightarrow P$ gives
\be
T^{ab}_{(2)}(q) \;\;=\;\;  -\frac{i}{2} \Delta \left( \frac{q_c J^{ac}q_d  J^{bd}}{q\cdot P} \right)   
 \;+\; \frac{im}{2} \int ds\, \left[ X^a X^b (q\cdot \ddot{X}) + 2 (q\cdot X) X^{(a}\ddot{X}^{b)} \right] .
\ee
Using \eqref{eq:divsubsub}, we can now finally say that
\be
iT_{(2)}^{ab}(-q) \;\;=\;\;  \Delta\mathscr{S}^{ab}_{(2)}(q, X, m\dot{X}) + (\textrm{terms that vanish when }{T^{ab}}_{,a} = 0),
\ee
thus identifying the sub-sub-leading soft factor localized to the ends of the worldline.

\subsection{Proof there are no further unambiguous boundary terms}

Now we have identified the leading, sub-leading, and sub-sub-leading soft factors, we can look at the general form of the stress tensor at $k$th order, i.e., the term
\be
T_{(k)}^{ab}(q) \;=\;   \frac{im}{k!} \int ds\, \left[ (iq\cdot X)^k \right] \frac{d}{ds} \left( \frac{\dot{X}^a\dot{X}^b}{q\cdot \dot{X}} \right)
\ee
Integrating by parts gives
\begin{align}
T_{(k)}^{ab}(q) \;\;&=\;\;   \frac{(i^{k+1})m}{k!} \int ds\, \left[ \frac{d}{ds} \left( (q\cdot X)^k  \frac{\dot{X}^a\dot{X}^b}{q\cdot \dot{X}} \right) -   \frac{d}{ds} \left( (q\cdot X)^k \right) \frac{\dot{X}^a\dot{X}^b}{q\cdot \dot{X}} \right] \\
&=\;\;   \frac{(i^{k+1})m}{k!} \int ds\, \left[ \frac{d}{ds} \left( (q\cdot X)^k  \frac{\dot{X}^a\dot{X}^b}{q\cdot \dot{X}} \right) -   k \left( (q\cdot X)^{k-1} \right)\dot{X}^a\dot{X}^b \right].
\end{align}
After some manipulation, the second ``bulk'' term here can be written as
\begin{align} \label{eq:proven_in_app}
-   k \left( (q\cdot X)^{k-1} \right)\dot{X}^a\dot{X}^b \;\;= & \;\; -\frac{k}{2}\frac{d}{ds}\left[(q\cdot X)^{k-1}\frac{d}{ds}(X^{a}X^{b})-X^{a}X^{b}\frac{d}{ds}(q\cdot X)^{k-1}\right] \nonumber \\
&\;\; + k(q\cdot X)^{k-1}X^{(a}\ddot{X}^{b)} - \frac{1}{2}k(k-1)X^{a}X^{b}(q\cdot X)^{k-2}(q\cdot\ddot{X}) \nonumber \\
& \;\; -\frac{1}{2}k(k-1)(k-2)X^{a}X^{b}(q\cdot \dot{X})^{2}(q\cdot X)^{k-3}.
\end{align}
From this, we see that the bulk term has one piece that localizes to the worldline endpoints (the first line), one piece that vanishes when the equations of motion are satisfied (second line, using eq. \eqref{eq:divk}), and then one further piece that is not localized on the boundary, and also does not vanish for any general reason (third line). Analogously to eq. \eqref{eq:jerror}, we see that this last piece vanishes explicitly for $k=0,1,2$. That is, it vanishes at leading, sub-leading, and sub-sub-leading order.

We can therefore conclude that for the point particle stress tensor, for small $|\mathbf{q}|$, we must have
\begin{align} \label{eq:Tbdry}
iT^{ab}(-q) \;\;= \,\, &\left[ \Delta \mathscr{S}^{ab}_{(0)}(q, m\dot{X}) + \Delta \mathscr{S}^{ab}_{(1)}(q, X, m\dot{X}) + \Delta \mathscr{S}^{ab}_{(2)}(q, X, m\dot{X}) \right] \nonumber \\
& \qquad + \; (\textrm{terms that vanish when }{T^{ab}}_{,a} = 0) + \mathcal{O}(|\mathbf{q}|^2),
\end{align}
and we have shown that at higher orders there are no terms which unambiguously localize to the future and past boundaries of the particle's worldline. This is the main result of the paper. It is what singles out the leading, sub-leading, and sub-sub-leading soft factors, and it explains the appearance of the sub-sub-leading term in \cite{soft}.

There are several lacunae in our arguments, which require more technical discussion -- this is given in  Appendix \ref{app.b}. There we explain how our classical reasoning applies to quantum scattering. That discussion encompasses scattering from $N$ incoming particles to $M$ outgoing ones, and we include hard gravitons. We demonstrate explicitly that the fact that these three terms are localized to the worldline boundary allows them to factor out of all tree level amplitudes in a simple theory containing only a scalar field coupled to gravitons. We also see how the contribution coming from terms that vanish when $T^{ab}_{,a} = 0$ must itself vanish at the quantum level, so that this contribution does not upset our result. In essence this contribution vanishes quantum mechanically because the equations of motion are satisfied at the operator level, or inside of a path integral.

\section{Discussion} \label{sec:done}

Our conclusions here are summarized in equations \eqref{eq:Jbdry} and \eqref{eq:Tbdry}. We have demonstrated that in a particular soft limit, the point particle sources $j^a$ and $T^{ab}$ contain contributions which are localized at the boundaries of the particle's worldline. These boundary contributions turn out to be simply related to the leading and sub-leading soft photon factors, and the leading, sub-leading, and sub-sub-leading soft graviton factors. They are precisely the terms which are known to factor out of tree level scattering amplitudes. Additionally, we showed that there are no further terms in a low frequency expansion which unambiguously localize to the boundary. This explains why tree level soft factorization stops after sub-leading order for photons and sub-sub-leading order for gravitons. We show this explicitly for a scalar field coupled to linearized gravity in appendix \ref{app.b}.

Looking forward, we also expect that our perspective may be able to shed more light on loop corrections to the soft theorems. In Appendix \ref{app.b} we have noted some places where such effects might be seen in our framework. It could also be illuminating to investigate the logarithmic corrections found in some of the work on ``classical soft theorems'' \cite{sahoo1,laddha0.1,laddha0.2,laddha1,saha,sahoo2,manu,mao}, using the approach presented here. Lastly, an extension of our results to non-minimal couplings to radiation fields could provide a foundation to further understand ``non-universal'' corrections to the soft factors beyond leading order \cite{laddha2,elvang1,bhatkar}. Such an extension could also include an analysis of spinning particles -- beyond leading order the soft factors which involve $J^{ab}$ are modified when the particles involved possess intrinsic angular momentum.


\appendix

\section{An example of soft factorization} \label{app.b}

In this appendix, we will demonstrate soft factorization at leading, sub-leading, and sub-sub-leading order, at tree level, in a theory consisting of a scalar field coupled to gravitons. We use a functional form of QFT without explicit reference to Feynman diagrams. However we will note which approximations we take that are equivalent to working at tree level diagrammatically.

First we will lay out our conventions and recall some basic results for scalar fields coupled to gravitons. We then discuss the simplest $2\rightarrow2$ scattering amplitude with an additional soft graviton emission, and see how soft factorization emerges. We then show how the factorization property generalizes simply to all hard amplitudes in the theory. Some of this has been discussed before by us, using the language of influence functionals \cite{us}.

\subsection{Computing scattering amplitudes using the generating functional}

To establish our notation and fix our conventions, we recall basic results for a free scalar field $\phi$ of mass $m$. The generating functional for the free field is
\be
\mathcal{Z}[J] \;\;\equiv\;\;  \int \mathcal{D} \phi \, \exp \left[ -i\int \phi K^{-1} \phi + i \int J \phi \right]  \;\;\; = \;\;\;   e^{\frac{i}{2}\int J KJ}
\ee
in which $ K^{-1}$ is shorthand for the free wave operator for the scalar field, $J$ is a source for $\phi$, and we use a compressed notation for integration, in which
\be
\int \phi K^{-1} \phi  \; \equiv \;  \int d^4x\, \phi(x) K^{-1}(x) \phi(x) ,
\ee
\be
\int J \phi \; \equiv \;  \int d^4x \, J(x) \phi(x),
\ee
and
\be
\int J KJ \; \equiv \;  \int d^4x \int d^4x' \, J(x) K(x,x') J(x').
\ee

To generate amplitudes from this generating functional, one uses the Lehmann-Symanzik-Zimmerman (LSZ) procedure \cite{IZ} for computing the usual scattering operator $S$ from a generating functional. For the free scalar field this is
\begin{equation}\label{eq:smatrix}
S \;\; =\;\; :e^{\int  \phi_{\textrm{in}} K^{-1}\frac{\delta}{\delta J}}:\,\mathcal{Z}[J]\bigg|_{J=0} .
\end{equation}
The colons here denote normal ordering, and the ``in'' field operator $\phi_{\textrm{in}}$ obeys the wave equation
\begin{equation}
K^{-1}(x) \, \phi_{\textrm{in}}(x) \;=\; 0,
\end{equation}
and is related to the full field $\phi$ via the weak asymptotic limits
\begin{equation}
\lim_{x^{0}\rightarrow-\infty}\big[\langle \beta|\phi(x)|\alpha\rangle-\langle \beta|\phi_{\textrm{in}}(x)|\alpha\rangle\big] \;=\; 0,
\end{equation}
in which $| \alpha \rangle, | \beta \rangle$ are arbitrary states of the system. We can split the scalar field into positive and negative frequency parts
\begin{equation}
\phi_{\textrm{in}}(x)= \phi^{+}_{\textrm{in}}(x)+\phi^{-}_{\textrm{in}} \;\;=\;\; \int d^{3}\mathbf{p} \; (\psi_{p}(x)a_{p} + h.c.\,),
 \label{phiIn}
\end{equation}
and demand the creation and annihilation operators obey the commutator $[a_{p},a^{\dagger}_{p'}]=(2\pi)^3\delta^{(3)}(\mathbf{p}-\mathbf{p}')$. If we normalize states as $|p\rangle\equiv a^{\dagger}_{p}|0\rangle$, the wavefunctions are normalized as
\begin{equation}
\psi_{p}(x) \;=\; \frac{e^{ip\cdot x}}{(2\pi)^{3}\sqrt{2E_{p}}},
 \label{psip}
\end{equation}
where $E_{p}=\sqrt{|\bm{p}|^{2}+m^{2}}$ is the energy of a particle.

The elements of the S-matrix give scattering amplitudes $A(\beta|\alpha)={}_{\textrm{in}}\langle\beta|S|\alpha\rangle_{\textrm{in}}$, and we are particularly interested in amplitudes of the form $A(\{p_m'\}|\{p_n\})$. To obtain these from the scattering operator in eq. \eqref{eq:smatrix}, we define
\begin{equation}
S_{(\{p_m'\}|\{p_n\})}[\delta_J] \;=\; \langle 0|a_{p'_{1}}\dots a_{p'_{M}}e^{\int \phi^{-}K^{-1}\frac{\delta}{\delta J}}e^{\int \phi^{+}K^{-1}\frac{\delta}{\delta J}}a^{\dagger}_{p_{1}}\dots a^{\dagger}_{p_{N}}|0\rangle,
\end{equation}
which gives scattering amplitudes after acting on $\mathcal{Z}$.
Commuting the creation/annihilation operators through the S-matrix we obtain the standard LSZ expression in terms of the amputation and on-shell restriction of the correlation function
\begin{align} \label{eq:amplitudes}
A(\{p_m'\}|\{p_n\})&= S_{(\{p_m'\}|\{p_n\})}[\delta_{J}]\mathcal{Z}[J]\bigg|_{J=0} \nonumber \\
&=\int d^{4}x'_{1}\dots d^{4}x'_{M}\,\bar{\psi}_{p'_{1}}(x'_{1})\dots \bar{\psi}_{p'_{M}}(x'_{M}) \nonumber \\
&\times\int d^{4}x_{1}\dots d^{4}x_{N}\,\psi_{p_{1}}(x_{1})\dots \psi_{p_{N}}(x_{N})  \\
&\times K^{-1}(x'_{1})\dots K^{-1}(x'_{M})K^{-1}(x_{1})\dots K^{-1}(x_{N}) \nonumber \\
&\times\frac{\delta}{\delta J(x'_{1})} \dots \frac{\delta}{\delta J(x'_{M})}  \frac{\delta}{\delta J(x_{1})} \dots\frac{\delta}{\delta J(x_{N})}\mathcal{Z}[J]\bigg|_{J=0}. \nonumber
\end{align}
From this expression one can compute any S-matrix element between these massive particle states.

If we now couple to gravitons in a linearized gravity theory, the generating functional becomes
\be\label{eq:zed0}
\mathcal{Z}[J, I] \equiv \int \mathcal{D} \phi \mathcal{D}h \, \exp \left[ -\frac{i}{2}\int \phi K^{-1} \phi + -\frac{i}{2}\int h D^{-1} h + i\kappa\int T[\phi]h + i \int J \phi + i \int I h \right]
\ee
in which $D^{-1}$ represents the free wave operator for the graviton field $\kappa h_{ab} \equiv g_{ab} - \eta_{ab}$, and $I^{ab}$ acts as a source for $h_{ab}$ (we have suppressed Lorentz indices here). $T^{ab}[\phi]$ is the stress tensor of the scalar, and $\kappa \equiv \sqrt{8\pi G}$. Doing the integrations, one gets
\be \label{eq:zed}
\mathcal{Z}[J, I] \;\;=\;\;  e^{\frac{1}{2} \sum_{j=1}^\infty \mathrm{Tr}  \left( - K \frac{-i\delta}{\delta I^{ab} }   \hat{T}^{ab}  \right)^j  } \;e^{\frac{i}{2}\int J K(\delta_I)J} \; e^{\frac{i}{2}\int I D I} .
\ee

Here $\hat{T}$ is a differential operator corresponding to the stress-energy tensor of the scalar, and $-i\delta/\delta I$ can be thought of as an insertion of the graviton field. A term involving the trace of $(K h_{ab}\hat{T}^{ab})^n$, in the first exponential factor, corresponds to a closed scalar loop with $n$ graviton insertions, and thus describes graviton polarization effects by scalar loops. We are not currently interested in looking at loop corrections, so we simply drop this factor in what follows, leaving
\be \label{eq:zedapprox}
\mathcal{Z}[J, I] \;\; \approx \;\;  e^{\frac{i}{2}\int J K(\delta_I)J} \; e^{\frac{i}{2}\int I D I}.
\ee
where in our compressed integral notation,
\be
\int I D I \;\;\equiv\;\; \int d^4x \int d^4x' \, I^{ab}(x) \; D_{abcd}(x,x') \; I^{cd}(x') , \;\;\;\;\;\textrm{etc.}
\ee

In \eqref{eq:zedapprox}, $K(\delta_I)$ is the propagator $K(x',x|h)$ for the scalar field in a ``frozen'' background $h_{ab}$, with the background field replaced by a functional derivative with respect to $I^{ab}$. This propagator can be written in path integral form (compare ref. \cite{peskinS}, problem 15.4):
\be \label{eq:PI}
K(x',x|h) \;=\;  i\int_0^\infty ds \int_{x}^{x'} \mathcal{D}X(s') \, e^{iS[X] |_0^s + i \kappa \int T^{ab}  h_{ab} |_0^s}
\ee
$T^{ab}$ here takes the form of the stress tensor for the relativistic point particle discussed in the main text, plus a contribution which is $\propto \eta^{ab}$. Note that the range on the proper time integrals in the exponent here is different than in \eqref{eq:EMcurrent} and \eqref{eq:Gcurrent}. This does not change our results in any way. This second contribution will vanish when contracted with the polarization tensor $\epsilon_{ab}$ of the soft graviton, so we can forget about it. A very important property of these path integrals is that the canonical momentum is not simply $m\dot{X}$. Instead,
\be \label{eq:canmom}
\hat{P} K(x',x|h) \;=\;  -i\partial_{x'}K(x',x|h) \;\;=\;\;  m\dot{x}' + \mathcal{O}(\kappa),
\ee
so the canonical momentum is altered due to the propagation in the background field. The same thing occurs when coupling a charged particle to a background vector potential $\vec{A}$ in non-relativistic quantum mechanics. The property \eqref{eq:canmom} will be important for what follows.

\subsection{$\mathbf{2\rightarrow 2}$ scattering, plus a soft graviton}

Using the LSZ formalism, we now generate the amplitude for a simple $2 \rightarrow 2$ scattering of scalar particles, along with the emission of a single graviton of momentum $q$ and polarization $\epsilon_{ab}$. The momentum of the graviton is assumed to be small. Generalizing eq. \eqref{eq:amplitudes} to include gravitons, we have:
\begin{align}
\mathscr{A}(p'_1, p'_2; q, \epsilon|p_1, p_2)&=\int d^{4}x'_{1} \int d^{4}x'_2\,\bar{\psi}_{p'_{1}}(x'_{1}) \bar{\psi}_{p'_2}(x'_2) \int d^4 z \, \bar{\Psi}_q(z) \bar{\epsilon}_{ab} \nonumber \\
&\times\int d^{4}x_{1}\int d^{4}x_{2}\,\psi_{p_{1}}(x_{1}) \psi_{p_2}(x_2) \\
&\times K^{-1}(x'_{1}) K^{-1}(x'_2) \left[ D^{-1}(z) \right]^{abcd} K^{-1}(x_{1}) K^{-1}(x_2) \nonumber \\
&\times\frac{\delta}{\delta J(x'_{1})} \frac{\delta}{\delta J(x'_2)}  \frac{\delta}{\delta I^{cd}(z)} \frac{\delta}{\delta J(x_{1})}\frac{\delta}{\delta J(x_2)}\mathcal{Z}[J,I]\bigg|_{J=I=0} \nonumber
\end{align}
Here, $\bar{\Psi}_q(z) \bar{\epsilon}_{ab}$ is the (complex conjugated) wavefunction of the emitted graviton, and we are acting on the approximate generating functional \eqref{eq:zedapprox}.
Performing the derivatives gives
\begin{align}
\mathscr{A}(p'_1, p'_2; q, \epsilon|p_1, p_2)&=\int d^{4}x'_{1} \int d^{4}x'_2\,\bar{\psi}_{p'_{1}}(x'_{1}) \bar{\psi}_{p'_2}(x'_2)  \nonumber \\
&\times\int d^{4}x_{1}\int d^{4}x_{2}\,\psi_{p_{1}}(x_{1}) \psi_{p_2}(x_2) \nonumber \\
&\times K^{-1}(x'_{1}) K^{-1}(x'_2)  K^{-1}(x_{1}) K^{-1}(x_2)  \\
&\times \left[ K(x'_1,x_1 | \delta_I)  K(x'_2,x_2 | \delta_I) + \mathrm{permutations} \right]  \nonumber  \\
&\times i \int d^4 z \, \bar{\Psi}_q(z) \, \bar{\epsilon}_{ab} \;  I^{ab}(z) \; e^{\frac{i}{2} \int I D I}\bigg|_{I=0}.\nonumber
\end{align}
To clean up our equations a bit, let us assume the particles are distinguishable, thereby removing the symmetrization in the fourth line above. This leaves
\begin{align}  \label{eq:amp_midway}
\mathscr{A}(p'_1, p'_2; q, \epsilon|p_1, p_2)&=\int d^{4}x'_{1} \int d^{4}x'_2\,\bar{\psi}_{p'_{1}}(x'_{1}) \bar{\psi}_{p'_2}(x'_2)  \nonumber \\
&\times\int d^{4}x_{1}\int d^{4}x_{2}\,\psi_{p_{1}}(x_{1}) \psi_{p_2}(x_2) \\
&\times K^{-1}(x'_{1}) K^{-1}(x'_2)  K^{-1}(x_{1}) K^{-1}(x_2) \nonumber \\
&\times K(x'_1,x_1 | \delta_I) \; K(x'_2,x_2 | \delta_I) \;\; i\int d^4 z \, \bar{\Psi}_q(z) \; \bar{\epsilon}_{ab} \;  I^{ab}(z) \; e^{\frac{i}{2}  \int I D I}\bigg|_{I=0}. \nonumber
\end{align}

Let us now pause and examine the expression
\be
 K(x'_1,x_1 | \delta_I) \; K(x'_2,x_2 | \delta_I) \;\;  i \int d^4 z \, \bar{\Psi}_q(z) \; \bar{\epsilon}_{ab} \;  I^{ab}(z) \; e^{\frac{i}{2} \; \int I D I}\bigg|_{I=0},
\ee
which appears in the last line of \eqref{eq:amp_midway}. The full amplitude \eqref{eq:amp_midway} is simply obtained by the usual LSZ prescription for the scalar field applied to this object.
By eq. \eqref{eq:PI}, this expression is equal to
\begin{align} \label{eq:e1}
- \int_0^\infty ds_1 \int_0^\infty ds_2 &\int_{x_1}^{x'_1} \mathcal{D}X_1\int_{x_2}^{x'_2} \mathcal{D}X_2 \nonumber \\
\times &\left[ e^{iS[X_1] |_0^{s_1} +iS[X_2] |_0^{s_2} } \; e^{i \kappa \int T_1^{ab}  \left( \frac{-i\delta}{\delta I^{ab}} \right)|_0^{s_1}  + i \kappa \int T_2^{ab}  \left( \frac{-i\delta}{\delta I^{ab}} \right) |_0^{s_2} }   \right] \nonumber \\
\times &i\int d^4 z \, \bar{\Psi}_q(z) \; \bar{\epsilon}_{ab} \;  I^{ab}(z) \; e^{\frac{i}{2} \int I D I}\bigg|_{I=0} 
\end{align}
and here the factors
\be
e^{i \kappa \int T^{ab}  \left( \frac{-i\delta}{\delta I^{ab}} \right)|_0^{s} }
\ee
act as linear shift operators on the functional of $I^{ab}$ on the last line of \eqref{eq:e1}, giving
\begin{align} \label{eq:e2}
- \int_0^\infty ds_1 \int_0^\infty ds_2 &\int_{x_1}^{x'_1} \mathcal{D}X_1 \int_{x_2}^{x'_2} \mathcal{D}X_2 \nonumber \\
&\times \left[ e^{iS[X_1, P_1] |_0^{s_1} +iS[X_2, P_2] |_0^{s_2}  + \frac{i\kappa^2}{2} \int  \left[T_1 + T_2 \right] D  \left[T_1 + T_2 \right] } \right] \nonumber \\
 &\times i \kappa \int d^4 z \, \bar{\Psi}_q(z) \; \bar{\epsilon}_{ab} \;  \left[T_1 + T_2 \right]^{ab}(z) .
\end{align}

There are several remarks to be made about this expression. First we see that the last line is ($\mathscr{N}_{ab}$ here is the same as in eq. \eqref{eq:Ngrav})
\begin{align} \label{eq:e3}
i\kappa\int d^4 z \, \bar{\Psi}_q(z) \; \bar{\epsilon}_{ab} \,  \left[T_1 + T_2 \right]^{ab}(z) &\;=\;\;  i\kappa\frac{\bar{\epsilon}_{ab}}{(2\pi)^3 \sqrt{|\mathbf{q}|}}  \left[T_1 + T_2 \right]^{ab}(-q) \nonumber \\  &\equiv \;\; i\kappa \mathscr{N}_{ab} \; \left[T_1 + T_2 \right]^{ab}(-q),
\end{align}
where $\mathscr{N}_{ab}$ here is the graviton wave-function defined in eq. \eqref{eq:Ngrav}; the external graviton gives an insertion of the momentum space stress tensor we studied in the main text. It even shows up in the form $iT(-q)$, just as it does in eq. \eqref{eq:Tbdry}.

Now note that because $q$ is on shell, and the emitted graviton momentum $|\mathbf{q}|$ is taken to be small, we will simply use our classical expression \eqref{eq:Tbdry} to write $i\left[T_1 + T_2 \right]^{ab}(-q)$ as a sum of boundary terms plus terms that vanish when ${\left[T_1 + T_2 \right]^{ab}}_{,a} = 0$.
For convenience we define just the boundary contribution
\be \label{eq:Tfrak}
i\mathfrak{T}_{1/2}^{ab}(-q) \equiv \,\,  \left[ \Delta \mathscr{S}^{ab}_{(0)}(q, m\dot{X}_{1/2}) + \Delta \mathscr{S}^{ab}_{(1)}(q, X_{1/2}, m\dot{X}_{1/2}) + \Delta \mathscr{S}^{ab}_{(2)}(q, X_{1/2}, m\dot{X}_{1/2}) \right],
\ee
and then write the term in \eqref{eq:e3} as
\be
i \kappa \mathscr{N}_{ab}\; \left[\mathfrak{T}_1 + \mathfrak{T}_2 \right]^{ab}(-q)
\ee
when $T_1 + T_2$ is conserved, and we have also neglected the terms higher order in $|\mathbf{q}|$.

We can see that the stress tensor $T_1 + T_2$ is indeed conserved within the path integrals of \eqref{eq:e2} as a consequence of the equations of motion. Note that the quantity
\be
\int_{x_1}^{x'_1}\mathcal{D}X_1 \int_{x_2}^{x'_2} \mathcal{D}X_2 \; e^{iS[X_1, P_1] |_0^{s_1} +iS[X_2, P_2] |_0^{s_2}  + \frac{i\kappa^2}{2} \int  \left[T_1 + T_2 \right] D  \left[T_1 + T_2 \right] }
\ee
can be rewritten (with appropriate gauge-fixing) as
\be
\int \mathcal{D}h \, e^{-i\int hD^{-1}h} \int_{x_1}^{x'_1}\mathcal{D}X_1 \int_{x_2}^{x'_2} \mathcal{D}X_2 \; e^{iS[X_1, P_1] |_0^{s_1} +iS[X_2, P_2] |_0^{s_2}  + i\kappa \int [T_1 + T_2] h}.
\ee
Because this expression is invariant under a change of bulk integration variables $h_{ab} + \delta h_{ab}$, $h_{ab}$ obeys the linearized Einstein equations sourced by $\left[T_1 + T_2 \right]^{ab}$ inside of the integral, which then implies that ${\left[T_1 + T_2 \right]^{ab}}_{,a} = 0$ in the integrand as well.

Finally, we note that the factor
\be
\exp \left(  \frac{i\kappa^2}{2} \int  \left[T_1 + T_2 \right] D  \left[T_1 + T_2 \right] \right)
\ee
in \eqref{eq:e2} is what contains the interaction between the two scalar particles mediated by virtual gravitons. We can now restrict our attention to the tree level contribution to this amplitude, which consists in making the replacement
\be \label{eq:tree_approx}
\exp \left(  \frac{i\kappa^2}{2} \int  \left[T_1 + T_2 \right] D  \left[T_1 + T_2 \right] \right)  \;\; \rightarrow \;\;  i\kappa^2 \int  T_1  D  T_2 .
\ee

Thus one virtual graviton is allowed to connect the two scalar particle lines in the Feynman diagram for this process, but we throw away terms where gravitons connect one line to itself, and terms with multiple gravitons connecting the two lines, as these diagrams all contain loops. The tree level amplitude should be $\mathcal{O}(\kappa^3)$ - it picks up a factor of $\kappa^2$ from the virtual graviton connecting the two scalar lines, and another factor of $\kappa$ from the vertex at which our soft graviton is emitted, so we have the graph:\\

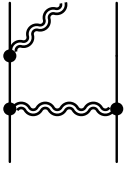
\be \label{tree22} \nonumber
\vspace{1em}
\begin{fmffile}{tree22}
\begin{fmfgraph*}(40,60)
\fmfstraight
\fmfleft{i1,o1} \fmfright{i2,o2} \fmftop{T}
\fmf{plain}{i1,vL,vT,o1}
\fmf{plain}{i2,vR,x,o2}
\fmffreeze
\fmf{dbl_wiggly}{vL,vR}
\fmf{dbl_wiggly}{vT,T}
\fmfdot{vL}
\fmfdot{vR}
\fmfdot{vT}
\end{fmfgraph*}
\end{fmffile}
\vspace{1em}
\ee
Henceforth we will drop all corrections that are not $\mathcal{O}(\kappa^3)$, to be sure we are isolating the tree level result.

The tree level contribution to \eqref{eq:e2} is thus
\begin{align} \label{eq:e4}
- \int_0^\infty ds_1 \int_0^\infty ds_2 &\int_{x_1}^{x'_1} \mathcal{D}X_1 \int_{x_2}^{x'_2} \mathcal{D}X_2 \\
\times &\left[ e^{iS[X_1, P_1] |_0^{s_1} +iS[X_2, P_2] |_0^{s_2} } \; i \kappa^2 \int  T_1  D  T_2  \right]  i\kappa \mathscr{N}_{ab} \; \left[\mathfrak{T}_1 + \mathfrak{T}_2 \right]^{ab}(-q) . \nonumber 
\end{align}
The boundary stress tensor insertions $\mathfrak{T}_{1/2}$ which appeared due to the soft graviton depend only on the boundary data $(q, x_{1/2}, x'_{1/2},m\dot{x}_{1/2},m\dot{x}'_{1/2})$. We cannot in general substitute dependence on e.g. $m\dot{x}_1$ with $-i \partial_{x_1}$, because of eq. \eqref{eq:canmom}. Importantly however, since the correction to this prescription for the conjugate momenta is higher order in $\kappa$, we can disregard the correction because it will not be part of the tree level result. This means we can factorize the boundary stress tensor out from the path integrals at this order, turning \eqref{eq:e4} into
\begin{align}   \label{eq:e5}
i \kappa \mathscr{N}_{ab} \; \left[\hat{\mathfrak{T}}_1 + \hat{\mathfrak{T}}_2 \right]^{ab}(-q) \int_0^\infty ds_1 \int_0^\infty ds_2 &\int_{x_1}^{x'_1} \mathcal{D}X_1 \int_{x_2}^{x'_2} \mathcal{D}X_2  \\
\times &\left[ e^{iS[X_1, P_1] |_0^{s_1} +iS[X_2, P_2] |_0^{s_2} } \; i \kappa^2 \int  T_1  D  T_2  \right] , \nonumber 
\end{align}
where now e.g. $\hat{\mathfrak{T}}_1$ is given by eq. \eqref{eq:Tfrak} with $m\dot{x}_1$ replaced simply with $-i \partial_{x_1}$, etc. Note that dropping the $\mathcal{O}(\kappa)$ corrections to the canonical momentum is equivalent to assuming that the particles travel with constant momentum in the infinite past and future, unaffected by long range gravitational forces between one another. Thus, e.g., at leading order in $1/|\mathbf{q}|$, and to tree level in $\kappa$, we have
\be \label{eq:factor1}
i\hat{\mathfrak{T}}_1^{ab}(-q) \;\;\approx \;\;  \frac{\hat{p}_1'^a\hat{p}_1'^b}{q\cdot\hat{p}_1'} - \frac{\hat{p}_1^a\hat{p}_1^b}{q\cdot\hat{p}_1}
\ee
which is a formal expression, in which
\be
\hat{p}_1'^a \equiv i\frac{\partial}{\partial x'_1}; \hspace{2em} \hat{p}_1^a \equiv -i\frac{\partial}{\partial x_1}.
\ee
Though a little tedious, one can check that in this form, the operators $\hat{\mathfrak{T}}_{1/2}$ also safely commute with the free field operators $K^{-1}$ in eq. \eqref{eq:amp_midway}, so we can factor them out of the entire amplitude by replacing $x_1 \rightarrow -i\partial_{p_1}$, $-i\partial_{x_1} \rightarrow p_1$, etc.

When we perform all of these replacements, eq. \eqref{eq:amp_midway} becomes
\begin{align}  \label{eq:amp_final}
\mathscr{A}(p'_1, p'_2; q, \epsilon|p_1, p_2)&=i\kappa\mathscr{N}_{ab} \left[\hat{\mathfrak{T}}_1 + \hat{\mathfrak{T}}_2 \right]^{ab}(-q)  \\
&\times \int d^{4}x'_{1} \int d^{4}x'_2\,\bar{\psi}_{p'_{1}}(x'_{1}) \bar{\psi}_{p'_2}(x'_2)\int d^{4}x_{1}\int d^{4}x_{2}\,\psi_{p_{1}}(x_{1}) \psi_{p_2}(x_2) \nonumber \\
&\times K^{-1}(x'_{1}) K^{-1}(x'_2)  K^{-1}(x_{1}) K^{-1}(x_2) \nonumber \\
&\times \int_0^\infty ds_1 \int_0^\infty ds_2 \int_{x_1}^{x'_1} \mathcal{D}X_1 \int_{x_2}^{x'_2} \mathcal{D}X_2\left[ e^{iS[X_1, P_1] |_0^{s_1} +iS[X_2, P_2] |_0^{s_2} } \; i\kappa^2 \int  T_1  D  T_2  \right]  \nonumber .
\end{align}

For concreteness, we will fully write out one of the stress tensor operators. We have (to sub-sub-leading order) that
\begin{align}
i\hat{\mathfrak{T}}_1^{ab}(-q) \;&=\;   \frac{p_1'^a p_1'^b}{q\cdot p_1'} - \frac{ p_1^a p_1^b}{q\cdot p_1}
+  i\left[  \frac{q_c \hat{J_1'}^{c(a} {p_1'}^{b)}}{q \cdot p'_1}    -    \frac{q_c \hat{J_1}^{c(a} p_1^{b)}}{q \cdot p_1}  \right]
-  \frac{1}{2}\left[  \frac{q_c \hat{J_1'}^{ac} q_d \hat{J_1'}^{bd} }{q \cdot p'_1}    -    \frac{q_c \hat{J_1}^{ac} q_d \hat{J_1}^{bd} }{q \cdot p_1} \right] \nonumber \\
&=\;  \sum_{k=0}^2 \left[  \mathscr{S}_{(k)}^{ab}(q,\hat{x'}_1,p'_1) -  \mathscr{S}_{(k)}^{ab}(q,\hat{x}_1,p_1) \right]
\end{align}
with $\hat{J_1'}^{ab} \equiv \left({p_1'}^a \hat{x'}_1^b - \hat{x'}_1^a {p_1'}^b \right)$,  $\hat{J_1}^{ab} \equiv \left({p_1}^a \hat{x}_1^b - \hat{x}_1^a {p_1}^b \right)$, $\hat{x'}_1 \equiv i \partial_{p'_1}$, and $\hat{x}_1 \equiv -i \partial_{p_1}$.
Finally, returning to eq. \eqref{eq:amp_final}, we find that for small $|\mathbf{q}|$,
\be
\mathscr{A}(p'_1, p'_2; q, \epsilon|p_1, p_2) \;\;\approx \;\;  i\kappa\mathscr{N}_{ab} \left[\hat{\mathfrak{T}}_1 + \hat{\mathfrak{T}}_2 \right]^{ab}(-q) A(p'_1, p'_2 |p_1, p_2),
\ee
where
\be
i\kappa\mathscr{N}_{ab} \left[\hat{\mathfrak{T}}_1 + \hat{\mathfrak{T}}_2 \right]^{ab}(-q)  \;\; = \;\;    \kappa \mathscr{N}_{ab} \sum_{k=0}^2 \left[\sum_{m=1}^2  \mathscr{S}_{(k)}^{ab}(q,\hat{x'}_m,p'_m) - \sum_{n=1}^2  \mathscr{S}_{(k)}^{ab}(q,\hat{x}_n,p_n) \right],
\ee
so we have agreement with the soft theorems up to sub-sub-leading order, at tree level.

\subsection{Generalization to all tree level scattering}

Having seen how soft factorization works in $2\rightarrow2$ scattering, it is relatively simple to show that the factorization holds also for all $N \rightarrow M$ scattering, both of scalars and of hard gravitons. Here we show what is involved in generalizing the $2\rightarrow2$ example to (i) $N \rightarrow N$ scalar particle scattering, (ii) $N \rightarrow M$ scalar particle scattering, and (iii) $N \rightarrow M$ scalar particle scattering with additional hard gravitons.

\paragraph{$\mathbf{N} \rightarrow \mathbf{N}$ scalar scattering\\}
Let us consider the contribution of \textit{connected} diagrams to the $N\rightarrow N$ amplitude (disconnected contributions themselves factor, so factorization extends trivially to those). Each connected diagram which contributes to $N\rightarrow N$ scattering at tree level takes the form
\be \label{treeNN}
\begin{centering}
\vspace{1em}
\begin{fmffile}{treeNN}
\begin{fmfgraph*}(60,60)
\fmfstraight
\fmfbottom{i1,B,i2, i3} \fmftop{o1,T,o2,o3} \fmfright{x1,R,x2,x3}
\fmf{plain}{i1,v1,vT,o1}
\fmf{plain}{i2,v2,vT2,o2}
\fmffreeze
\fmf{dbl_wiggly}{vT,T}
\fmf{dbl_wiggly}{v1,v2}
\fmf{dbl_wiggly}{v2,R}
\fmfdot{v1}
\fmfdot{v2}
\fmfdot{vT}
\end{fmfgraph*}
\end{fmffile}
\dots
\begin{fmffile}{treeNN2}
\begin{fmfgraph*}(60,60)
\fmfstraight
 \fmfbottom{i1,i2, i3} \fmftop{o1,o2,o3} \fmfleft{x1,L,x2,x3} \fmfright{i3,o3}
\fmf{plain}{i2,v2,vT2,o2}
\fmf{plain}{i3,v3,vT3,o3}
\fmffreeze
\fmf{dbl_wiggly}{v2,v3}
\fmf{dbl_wiggly}{L,v2}
\fmfdot{v2}
\fmfdot{v3}
\end{fmfgraph*}
\end{fmffile}
\end{centering} \hspace{1em},
\ee
or a permutation of this. We will always have $N$ scalar lines connecting the in/out particles, which are themselves connected by $N-1$ graviton lines, while the external soft graviton is attached to one of the incoming or outgoing scalar lines. As before, each internal graviton propagator contributes a factor of $\kappa^2$, and the vertex where the soft graviton is emitted contributes a factor of $\kappa$. This makes the entire diagram $\propto \kappa^{2N-1}$.

In terms of our framework, having $N$ scalar lines means eq. \eqref{eq:amp_midway} will be altered to include $N$ factors of $K(x',x|\delta_I)$. So we will have
\be
K(x_1',x_1|\delta_I) \cdots K(x_N',x_N|\delta_I) \; + \;  \textrm{permutations},
\ee
along with their associated amputations. If we restrict our results to tree level, the analog of eq. \eqref{eq:tree_approx} becomes
\be
\exp \left(  \frac{i\kappa^2}{2} \int  \left[\sum_{n=1}^N T_n \right] D  \left[\sum_{n=1}^N T_n \right] \right) \; \rightarrow \;  \left( i\kappa^2 \int  T_1  D  T_2 \right) \times \cdots \times \left( i\kappa^2 \int  T_{N-1}  D  T_N \right),
\ee
plus permutations. This inserts the $N-1$ graviton lines in \eqref{treeNN}, and is $\mathcal{O}(\kappa^{2N-2})$. The soft graviton vertex is $\mathcal{O}(\kappa)$, so once we include it, we have ``used up'' all of factors of $\kappa$ that we expect to appear in \eqref{treeNN}. This again leads to the same power-counting argument we used to go from \eqref{eq:e4} to \eqref{eq:e5}, and allows us to very simply factorize any occurrence of the boundary stress tensor out of the path integrals, and then the entire amplitude, at tree level.

The simple relationship between our explicit $2\rightarrow2$ example and the more general $N\rightarrow N$ case thus allows us to conclude that tree level soft factorization holds to sub-sub-leading order for all $N\rightarrow N$ scalar scattering in this theory.

\paragraph{$\mathbf{N} \rightarrow \mathbf{M}$ scalar scattering\\}
The even more general $N\rightarrow M$ scattering case is only marginally more difficult to handle than the $N\rightarrow N$ case. First let us think about what diagrams we expect to see in these amplitudes at tree level. Because the linearized theory only contains the linearized graviton-matter vertex, 
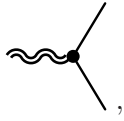
\be \label{gravitonvertex} \nonumber
\begin{fmffile}{gravitonvertex}
\begin{fmfgraph*}(40,40)
\fmfleft{i1} \fmfright{o1,o2}
\fmf{dbl_wiggly}{i1,v}
\fmf{plain}{o1,v,o2}
\fmfdot{v}
\end{fmfgraph*}
\end{fmffile},
\ee
each external scalar line must be connected to another at tree level (meaning $N+M$ must be even). Connected diagrams which contribute at tree level will thus have a similar structure as in the $N\rightarrow N$ case. There will be $(\frac{N+M}{2})$ scalar lines connected by $(\frac{N+M}{2}) - 1$ virtual graviton lines, and the power-counting arguments we have been using still hold.

We will use an example of this kind of scattering to highlight one further subtlety. The $2 \rightarrow 4$ amplitude contains the contribution \\

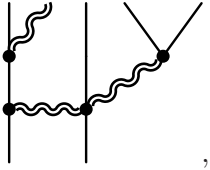
\be \label{treeNM} \nonumber
\begin{centering}
\vspace{1em}
\begin{fmffile}{treeNM}
\begin{fmfgraph*}(72,60)
\fmfstraight
\fmfbottom{i1,B,i2, i3,i4,i5} \fmftop{o1,T,o2,o3,o4,o5}
\fmf{plain}{i1,v1,vT,o1}
\fmf{plain}{i2,v2,vT2,o2}
\fmf{phantom}{B,vC,vTC,T}
\fmf{phantom}{i3,v3,vT3,o3}
\fmf{phantom}{i4,v4,vT4,o4}
\fmf{phantom}{i5,v5,vT5,o5}
\fmffreeze
\fmf{plain}{o3,vT4,o5}
\fmf{dbl_wiggly}{vT,T}
\fmf{dbl_wiggly, tension=0}{v1,v2}
\fmf{dbl_wiggly}{v2,vT4}
\fmfdot{v1}
\fmfdot{v2}
\fmfdot{vT}
\fmfdot{vT4}
\end{fmfgraph*}
\end{fmffile}
,
\end{centering}
\ee
which contains a scalar line connecting two of the outgoing particles. The only thing this changes in our formalism is that the path integral propagators $K(x',x|\delta_I)$ now occur with $x, x'$ corresponding to two incoming or outgoing particles, rather than one of each.

Because we have represented the soft factors as (future boundary term $-$ past boundary term) in the stress tensor, one might be concerned that the path integral propagator between e.g. two outgoing particles will induce a relative minus sign between their soft factors (as in eq. \eqref{eq:factor1}), contradicting the soft theorems. This seems to happen initially when factorizing the boundary stress tensor out of the path integral. However as long as one is careful when further factorizing the terms outside of the entire amplitude (see the lead up to eq. \eqref{eq:amp_final}), one sees that the LSZ procedure indeed provides the correct sign for every term.

Soft factorization to sub-sub-leading order then holds at tree level for $N\rightarrow M$ scalar scattering.

\paragraph{Adding hard gravitons\\}
The last and most general kind of amplitude we can consider in this theory is an $N\rightarrow M$ scalar amplitude which also contains any number of incoming or outgoing hard gravitons, in addition to our single emitted soft graviton. It is very straightforward to see that adding these gravitons will not upset factorization.

Consider adding a (hard) outgoing graviton of momentum $k$ and polarization $E$ to the out state of our $2\rightarrow2$ example. The last line of eq. \eqref{eq:amp_midway} becomes
\begin{align}
 K(x'_1,x_1 | \delta_I)  &K(x'_2,x_2 | \delta_I)   \\
 \times &\int d^4 z' \, \bar{\Psi}_k(z') \bar{E}_{cd} \left[ D^{-1}(z') \right]^{cdef} \frac{\delta}{\delta I^{ef}(z')} \, i\int d^4 z \, \bar{\Psi}_q(z) \, \bar{\epsilon}_{ab} \,  I^{ab}(z) \; e^{\frac{i}{2} \int I D I}\bigg|_{I=0} \nonumber .
\end{align}
Now, if the hard graviton functional derivative $\delta/\delta I^{ef}(z')$ hits the soft graviton contribution
\be
 i\int d^4 z \, \bar{\Psi}_q(z)  \; \bar{\epsilon}_{ab} \,  I^{ab}(z) ,
\ee
it will produce a term which vanishes, because
\be
D^{-1}(z')\bar{\Psi}_q(z') = 0.
\ee
The upshot is that the hard graviton contribution then just harmlessly commutes past the soft graviton contribution. This is true for any number of incoming or outgoing hard gravitons, and so our factorization arguments are still unaffected. Soft factorization occurs up to sub-sub-leading order for all tree level amplitudes in this theory.

\subsection{Conclusions}

What have we learned from this? We have seen that when using the path integral representation of the scalar propagator on a graviton background, eq. \eqref{eq:PI}, the identification of the soft factors as boundary terms in the stress tensor shows itself directly in the calculation of soft-graviton-added amplitudes. This is particularly simple at tree level, where we have ignored polarization of gravitons due to scalar loops. Additionally, propagation on a graviton background changes the canonical momentum of the scalar in the path integral (eq. \eqref{eq:canmom}), but power counting arguments allowed us to essentially ignore this subtlety at tree level, and immediately factorize the boundary terms.

So we have learned that the classical intuition presented in the main text carries over rather simply to all scalar-graviton amplitudes at tree level. It is known, however, that beyond leading order, the soft theorems acquire loop corrections. Importantly, we can look back at the $2\rightarrow2$ example to identify two places where loop corrections could arise in our framework. In particular, one could reintroduce graviton polarization effects to the generating functional, using \eqref{eq:zed} rather than \eqref{eq:zedapprox}. One could also keep track of corrections to the canonical momentum at higher order in $\kappa$ when attempting to factor the boundary stress tensor out of the path integral propagators. This would yield corrections to eq. \eqref{eq:e5}.



\end{document}